\documentclass[aps,pre,twocolumn,float]{revtex4}
\usepackage{amsmath,bm,epsfig}
\newcommand{\B}[1]{{\bm{#1}}}
\begin{document}

\title{The yield of Amorphous Solids Under Stress Control at Low Temperatures}
\author{ Valery Ilyin,  Itamar Procaccia, Carmel Shor and Murari Singh}
\affiliation{Dept. of Chemical Physics, The Weizmann Institute of Science,  Rehovot 76100, Israel}
\begin{abstract}
The yield of amorphous solids like metallic glasses under external stress was discussed asserting that it is related to the glass transition by increasing temperature, or that it can be understood using statistical theories of various sorts. Here we study the approach to stress-controlled yield and argue that neither assertions can be supported, at least at low temperatures. The yield of amorphous solids at low temperatures is a highly structured phenomenon, characterized by a specific series of mechanical instabilities, and having no similarity at all to fluidization by
increased temperature, real or fictive. The series of instabilities followed by stress controlled yield at low but
finite temperature protocols can be predicted by analyzing athermal quasi-static strain controlled protocols,
making the latter highly relevant for the deep understanding of the mechanical properties of amorphous solids.
\end{abstract}
\maketitle

The lack of crystalline order in amorphous solids makes the study of their mechanical properties a very
challenging subject of high current interest. One particularly intriguing characteristic is their yield-stress $\sigma_{\rm Y}$
which is defined as the lowest external mechanical stress above which unbounded plastic flow can be reached \cite{09RS,11Voi}. For
lower values of the external stress the amorphous solid deforms but remains able to develop an internal stress
that balances the external stress, leaving the system in mechanical equilibrium. Naturally one wants to understand
the yielding process, sometime with the hope of learning how to improve the material
atomic composition or its protocol of preparation. In this Letter we address the atomistic theory of the {\em approach} to yield under
stress control protocols.

Scanning the literature on the dynamics of yielding of amorphous solids one finds basically two
schools of thought. One school attempts to describe yielding using one or another sort of statistical theory that
assumes that the process can be modeled as a random series of creation and annihilation of soft spots, or regions
of plasticity that are sometime referred to as shear transformation zones \cite{12LE}. Another school of thought asserts that
mechanical yielding under stress can be understood as a kind of fluidization, similar to the inverse glass transition due to the increase
of temperature \cite{10GCE}. Both schools of thought may or may not invoke a notion of fictive or effective temperature which,
in order to explain the yield,
is assumed to increase even though the thermodynamic temperature remains low \cite{07HL}. We will provide strong evidence in this Letter that yielding is a highly structured, not random at all, series of instabilities that can be predicted
a-priori from zero-temperature considerations. Moreover, we will show that that there is no increase in fictive temperature near yield, and there is no necessity to invoke such an increase. The phenomenon of yield is purely mechanical and is easy to understand
once the right theoretical framework is provided.

To understand the phenomenon of stress controlled yield one requires a good method to observe the phenomenon with a
high degree of accuracy. In recent work (where full details can be found) such a simulational method was described using a ``variable shape Monte-Carlo" technique \cite{15DIMP}.
This method introduces strain into the simulation box by first defining a square box of {\em unit area} where the particles are at positions $\B s_i$. Next one defines a linear transformation $\B h$, taking the particles to positions $\B r_i$ via $\B r_i={\bf h}\cdot\B s_i$. In order to prevent rotations of the simulation box, the matrix ${\bf h}$ should be symmetric. The current area of a system becomes the determinant
$V=\mid {\bf h}\mid$. To proceed, two kinds of trial moves are considered: in the first we perform $n$ standard Monte Carlo moves (displacement of the particle positions given by $\vec{s}_i$)
\begin{equation}
\B s^{new}_i=\B s^{old}_i+\delta \B s,\hspace{4 mm} 1\le i\le N.
\label{Rmove}
\end{equation}
In this equation the $\alpha$ component of the displacement vector of a particle is given by
\begin{equation}
\delta s^{\alpha}=\Delta s_{max}(2\xi^\alpha-1),
\label{ParDisp}
\end{equation}
where $\Delta s_{max}$ is the maximum displacement and
$\xi^\alpha$ is an independent random number uniformly distributed between 0 and 1.
After $n$ sweeps defined by Eq.~(\ref{Rmove})
the transformation ${\bf h}$ changes according to
\begin{equation}
\B h^{new}=\B h^{old}+\delta\B h,
\label{hnew}
\end{equation}
where elements of the random symmetric matrix $\delta\B h$ are defined by
\begin{equation}
\delta h_{ij}=\Delta h_{max}(2\xi_{ij}-1),\hspace{4mm} i\le j.
\label{transH}
\end{equation}
Here $\Delta h_{max}$ is the maximum allowed change of a matrix element and
$\xi_{ij}$ is an independent random number uniformly distributed between 0 and 1.
The value of $\Delta h_{max}$  and the maximum displacement of particle
positions $\Delta s_{max}$ are selected to obtain a desired acceptance rate (usually between $30\%$ and $50\%$).
For each kind of move the trial configuration is accepted with probability
\begin{equation}
P_{tr}=\min \bigg[1,\exp\bigg(-\frac{\Delta G}{T}\bigg)\bigg].
\label{trial}
\end{equation}
Here $\Delta G$ is the {\em enthalpy change} due to the move,
\begin{equation}
\Delta G^{\prime}=U(\B \gamma+\B {\delta \gamma},\B r_i^{new})-
U(\B \gamma,\B r_i^{old})-V\sigma^{ext}_{xy}\delta\gamma \ .
\label{entG}
\end{equation}
where $\gamma$ is the system's strain. Temperature is measured here in Lennard-Jones units with the Boltzmann constant taken as unity.
To specialize the technique to stress-controlled simple shear simulations at
any desired temperature one chooses the
following ${\bf h}$ matrix
\begin{equation}
{\bf h}=L\left(
\begin{array}{c c}
1&\gamma \\
0&1
\end{array}
\right),
\label{hG}
\end{equation}
where $L$ is the length of the square simulation box and $\gamma$ is the simple shear strain, with the volume of the system V=$L^2$ being conserved. The external stress in this protocol is given by
\begin{equation}
\B \sigma^{ext}=\left(
\begin{array}{c c}
0&\sigma^{ext}_{xy} \\
\sigma^{ext}_{xy}&0
\end{array}
\right) \ .
\label{ExtStr}
\end{equation}
\begin{figure}
\epsfig{width=.23\textwidth,file=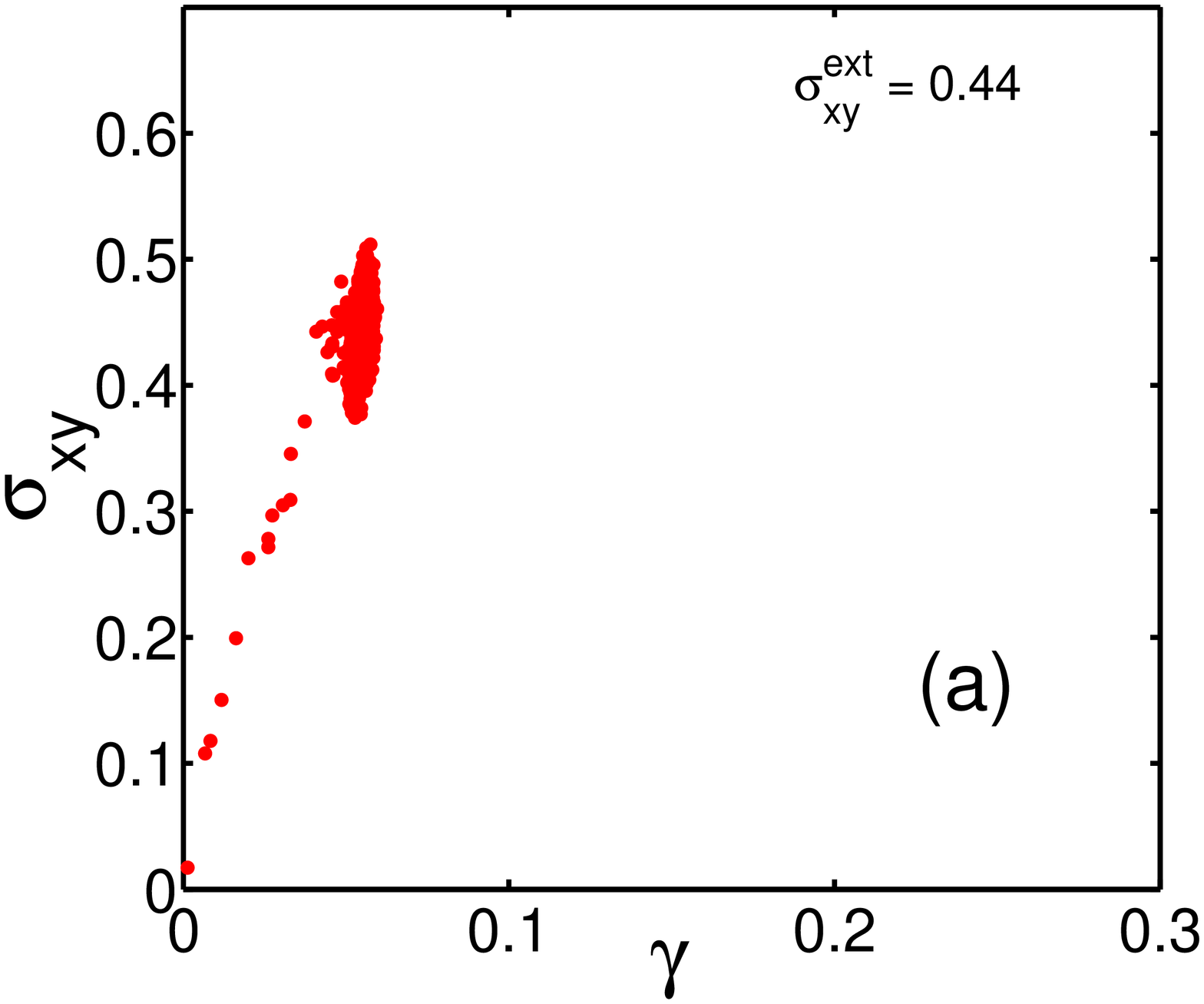}
\epsfig{width=.23\textwidth,file=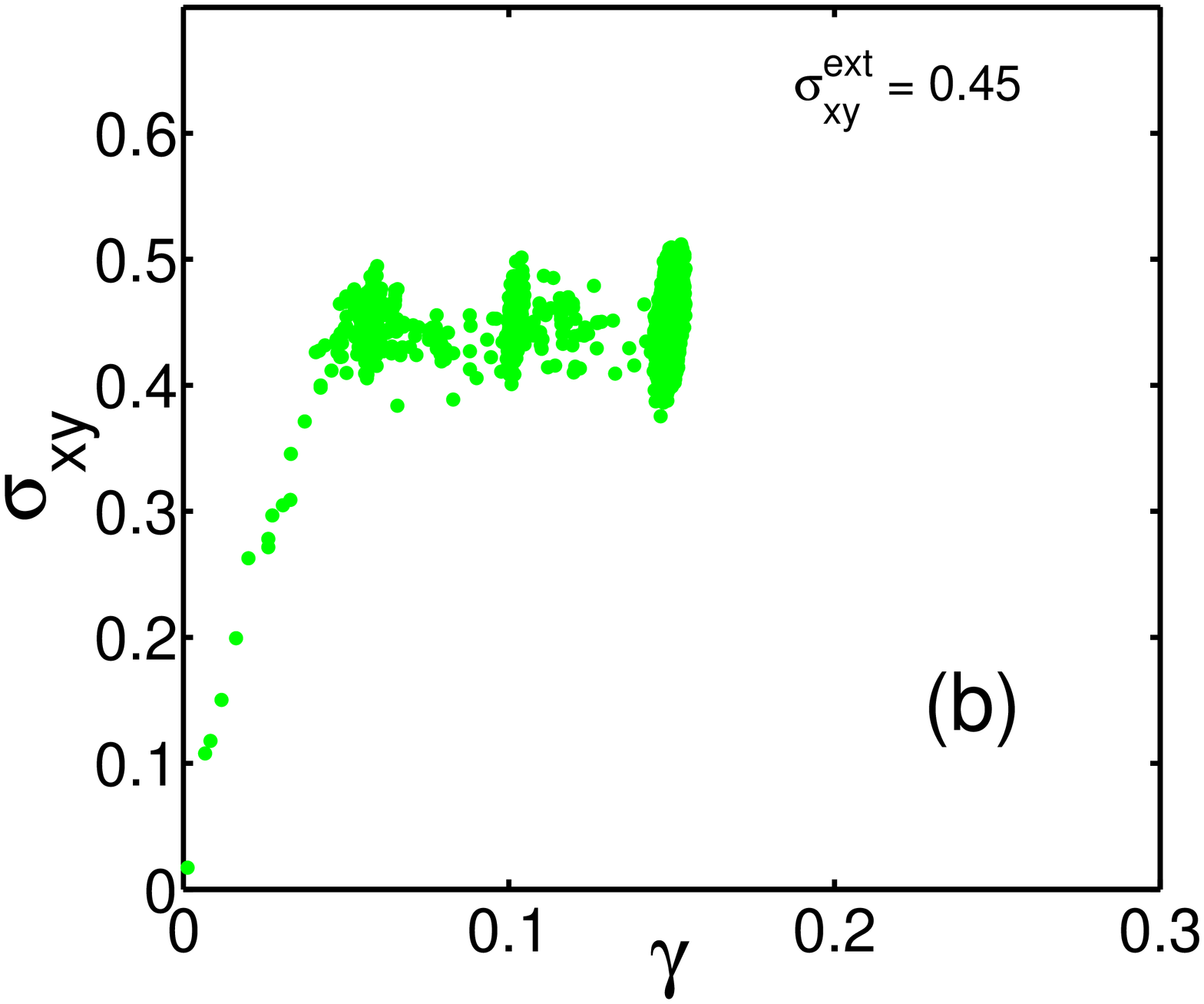}
\epsfig{width=.23\textwidth,file=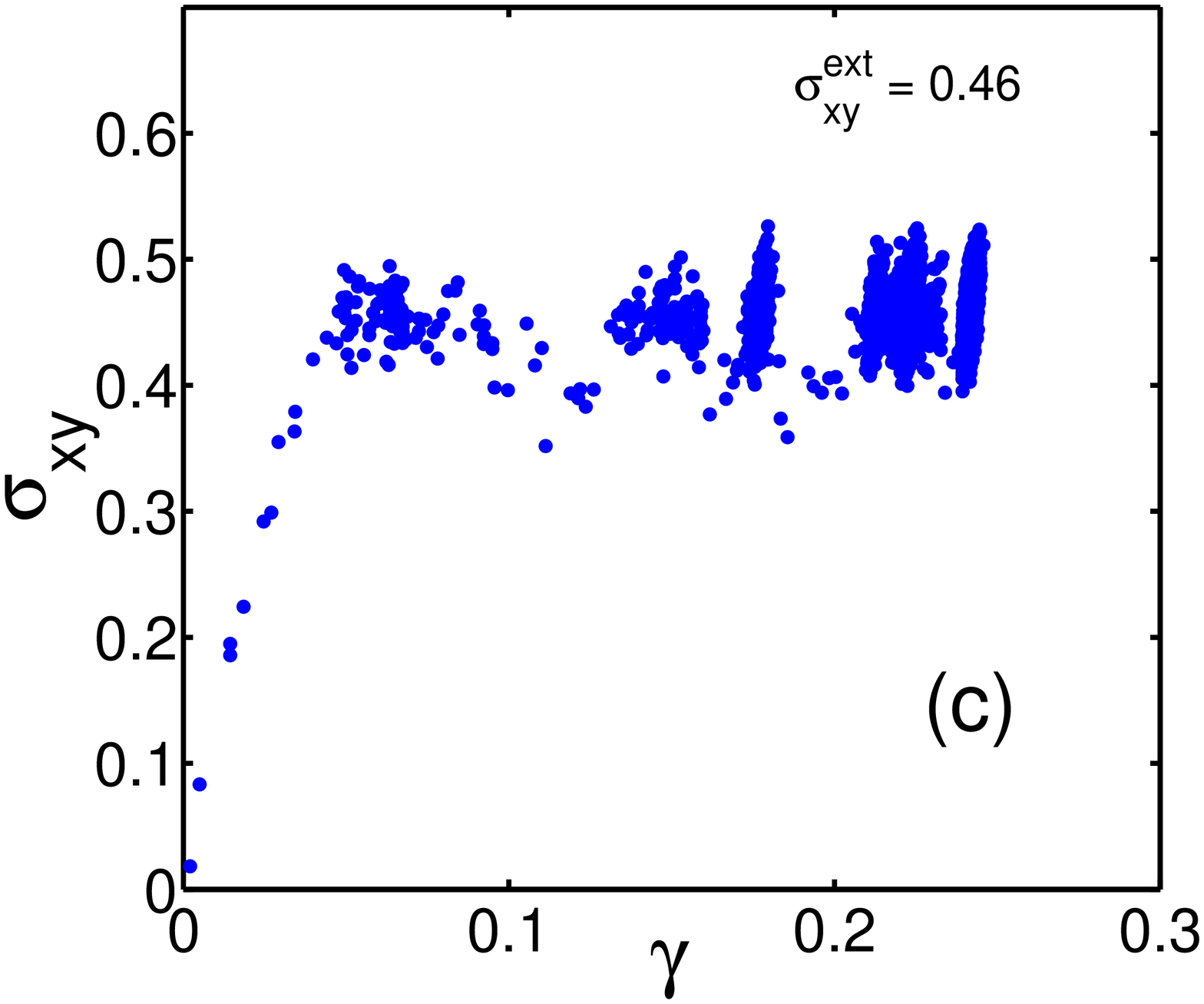}
\epsfig{width=.23\textwidth,file=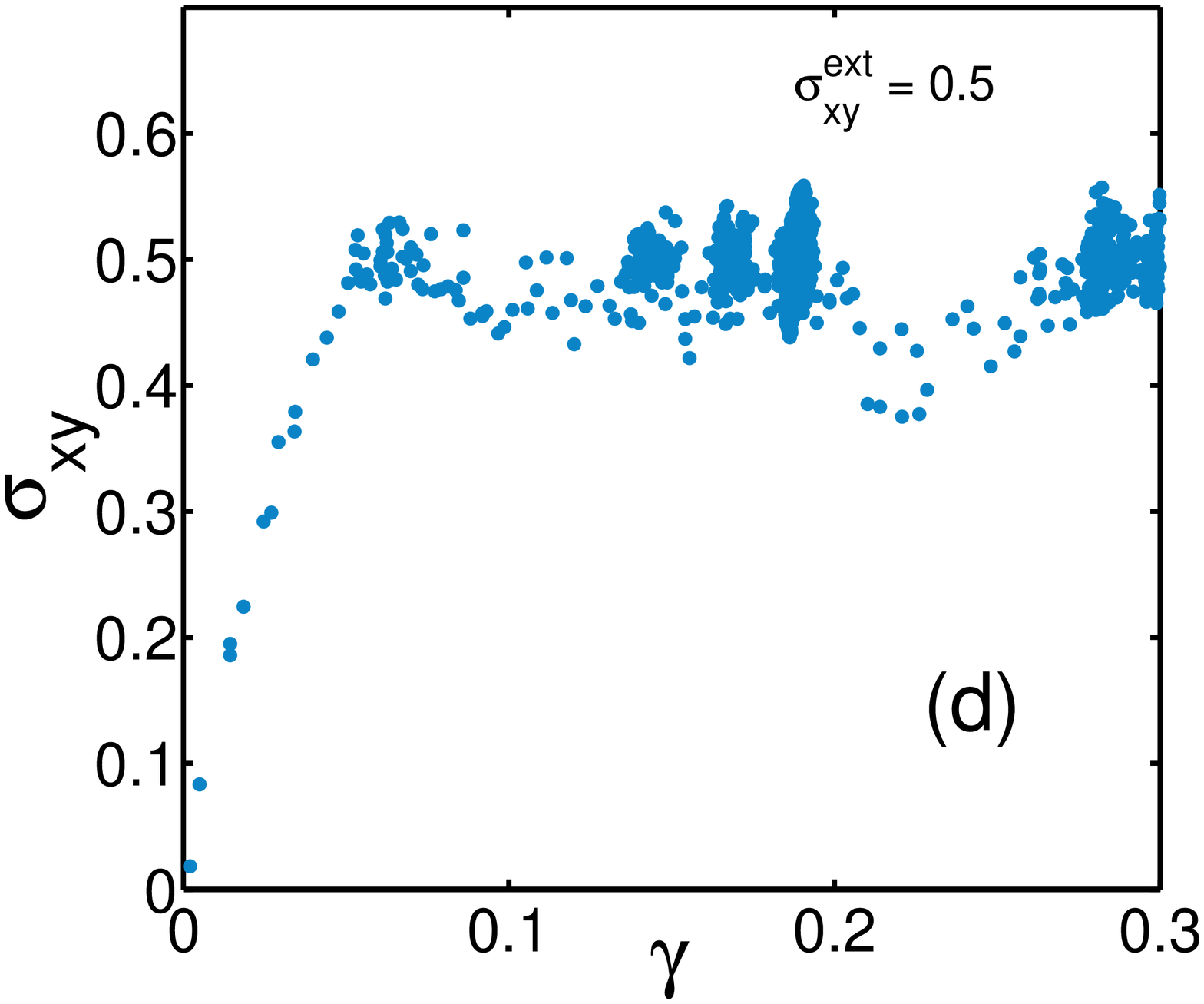}
\caption{Internal stress $\sigma_{xy}$ vs. strain $\gamma$ in one realization for four different values of external stress $\sigma^{ext}_{xy}$
at a temperature $T=0.005$ in Lennard Jones units. Panel a-d: $\sigma^{ext}_{xy}=0.44,0.45,0.46, 0.5$ respectively.}
\label{fig1}
\end{figure}

The nature of the results of the Monte Carlo protocol is presented in Fig~\ref{fig1}. In panel (a) the external stress is
$\sigma^{ext}_{xy}=0.44$ and the system equilibrates with stress fluctuations around the value of the external stress.
The red cloud of points represents the values of $\gamma$ and $\sigma_{xy}$ of the configurations that are visited during  the Monte Carlo
process. The system reaches detailed balance under the action of the enthalpy [ Eq.~(\ref{entG})].  Panels (b) and (c) show that for
slightly higher external stress $\sigma^{ext}_{xy}=0.45$ and $\sigma^{ext}_{xy}=0.46$ the system visits several metastable
states before arresting again at values of the internal stress that fluctuate around the external stress. When the system
arrests it again reaches detailed balance. (Sometime during the equilibration process the system resides in configurations that are part of  a metastable state long enough
to reach ``almost-detailed-balance" also there, before discovering the escape channel). Panel (d) shows that for $\sigma^{ext}_{xy}=0.5$
the system collapses mechanically and distorts indefinitely with $\gamma$ increasing without any obvious limit. Thus
we can conclude that for the present realization at a temperature of $T=0.005$ the yield stress lies between $0.46\le \sigma_{{Y}}\le 0.5$.

\begin{figure}
\vskip 0.5 cm
\epsfig{width=.32\textwidth,file=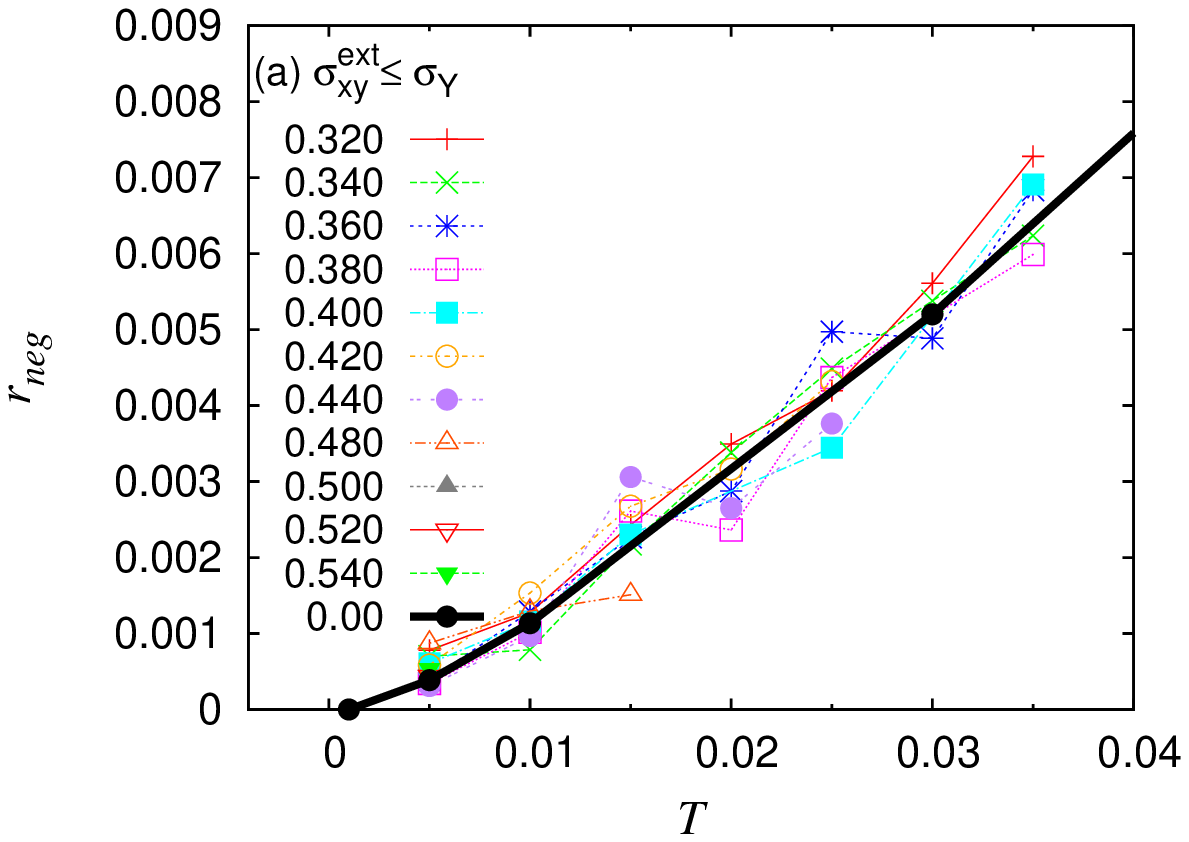}
\epsfig{width=.32\textwidth,file=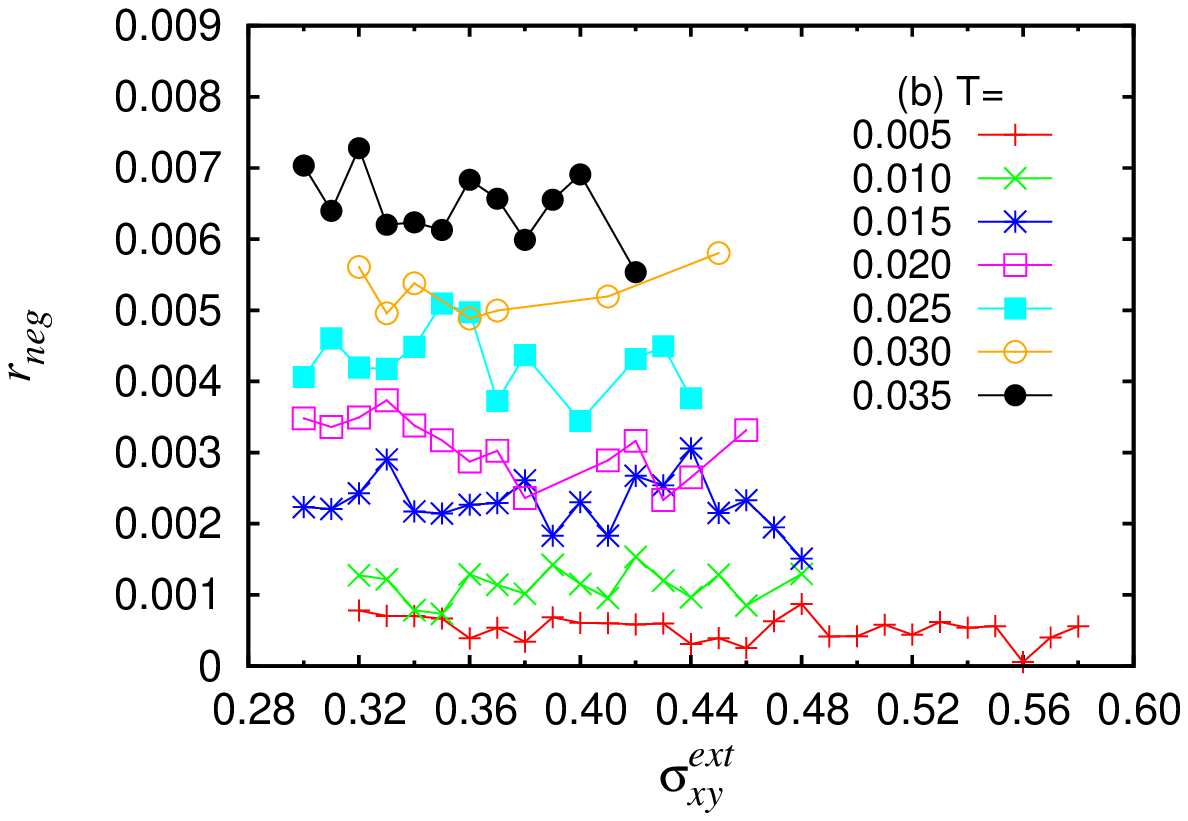}
\caption{Average over 500 realizations of the  dependence of the fraction of negative eigenvalues of Hessian matrix $r_{\rm neg}$
as a function of stress
and temperature. Upper panel: different external stresses as a function of temperature. Lower panel:
different temperatures as a function of external stress.}
\label{fig2}
\end{figure}

At this point we can ask whether approaching the yield point from below is similar in any way to approaching the glass transition point by
increasing the temperature. To this aim we study the Hessian matrix which is defined as
\begin{equation}
H_{i,j}^{\alpha,\beta} \equiv \frac{\partial^2 U(\B r_1,\B r_2,\cdots \B r_N)}{\partial r_i^\alpha \partial r_j^\beta} \ .
\end{equation}
At mechanical equilibrium and $T=0$, The Hessian matrix has $2N$ eigenvalues that are all positive.
At finite temperature one needs to arrest a given configuration and compute the Hessian $\B H$ for the arrested positions. Then the spectrum contains negative
eigenvalues whose number $n_{\rm neg}$ increases with the temperature \cite{85LS,02GCGP}. If the approach to yield were analogous to the increase
in temperature towards the glass transition, one would expect $n_{\rm neg}$ to increase as we approach a yield point. In our simulations we can arrest the Monte Carlo protocol at any value
of $\sigma^{ext}_{xy}$ and any value of T for which the system reaches detailed balance with $\langle\sigma_{xy}\rangle =\sigma^{ext}_{xy}$.
We then compute the Hessian matrix for these arrested configurations with the results shown in Fig.~\ref{fig2}. We find that the ratio
of the number of negative eigenvalues to $2N$ is fairly independent of $N$ and we thus report the number $r_{\rm neg}\equiv n_{\rm neg}/2N$.
In the upper panel we present the average over 500 realizations of the ratio of negative eigenvalues $r_{\rm neg}$ for zero external stress (black line) and for various external stresses up to the yield stress as a function of temperature. One sees that $r_{\rm neg}$ does not depend at all
on the external stress. The system does not ``prepare itself" for failure by increasing a fictive temperature. In the lower
panel the same conclusion is drawn from the independence of $r_{\rm neg}$ on the external stress, here shown for different
temperatures.

\begin{figure}
\vskip 0.5 cm
\epsfig{width=.23\textwidth,file=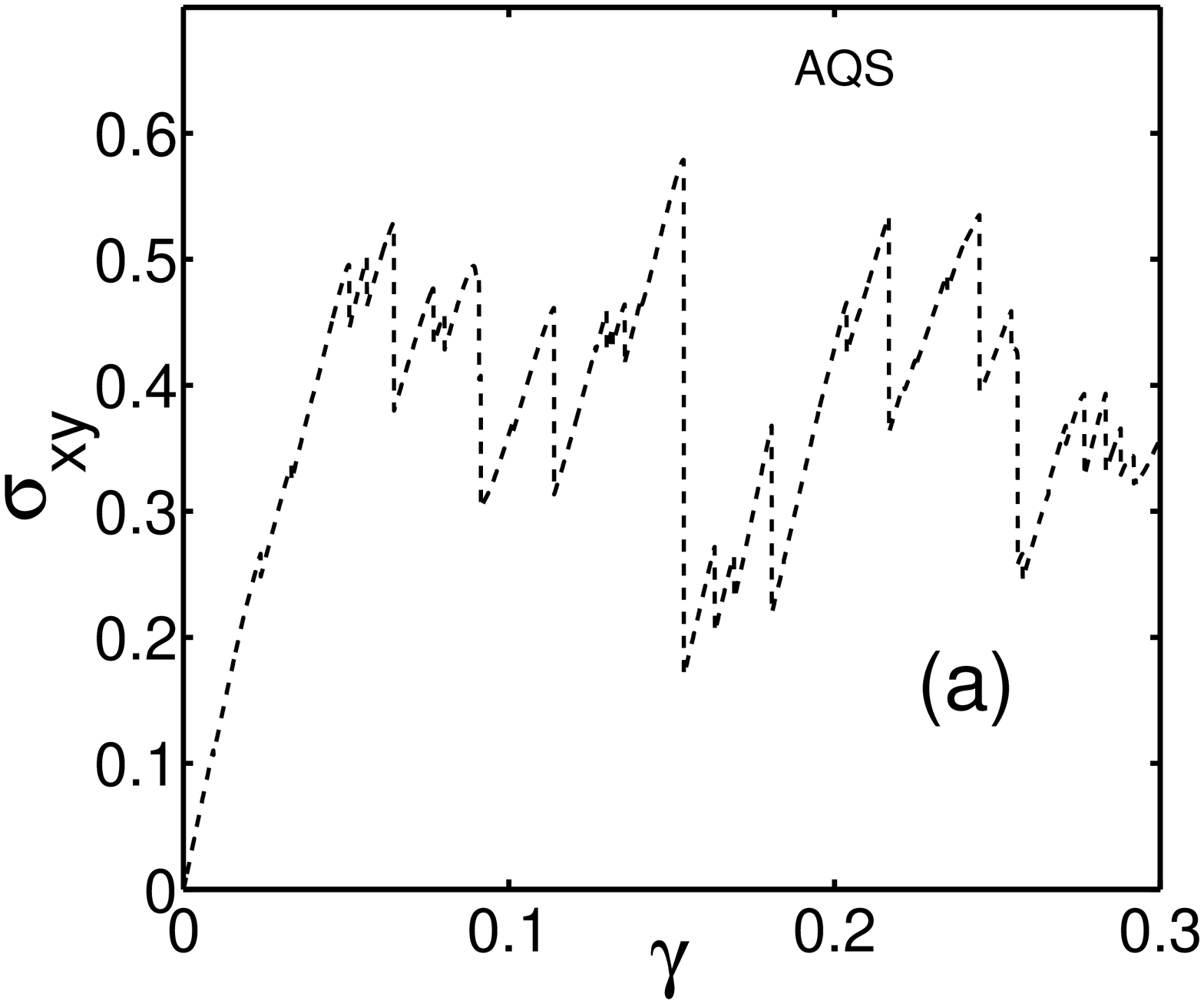} \\
\epsfig{width=.23\textwidth,file=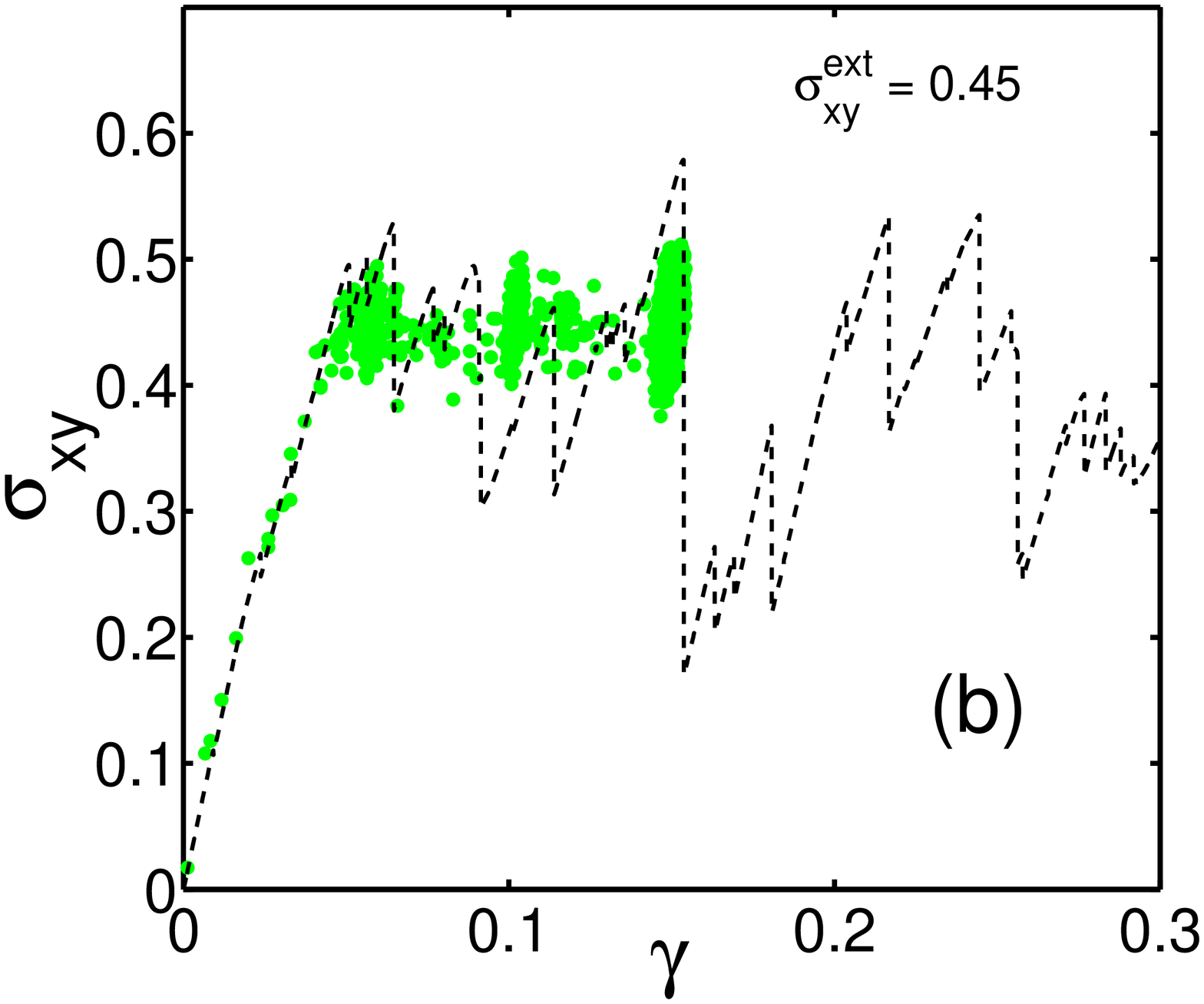}
\epsfig{width=.23\textwidth,file=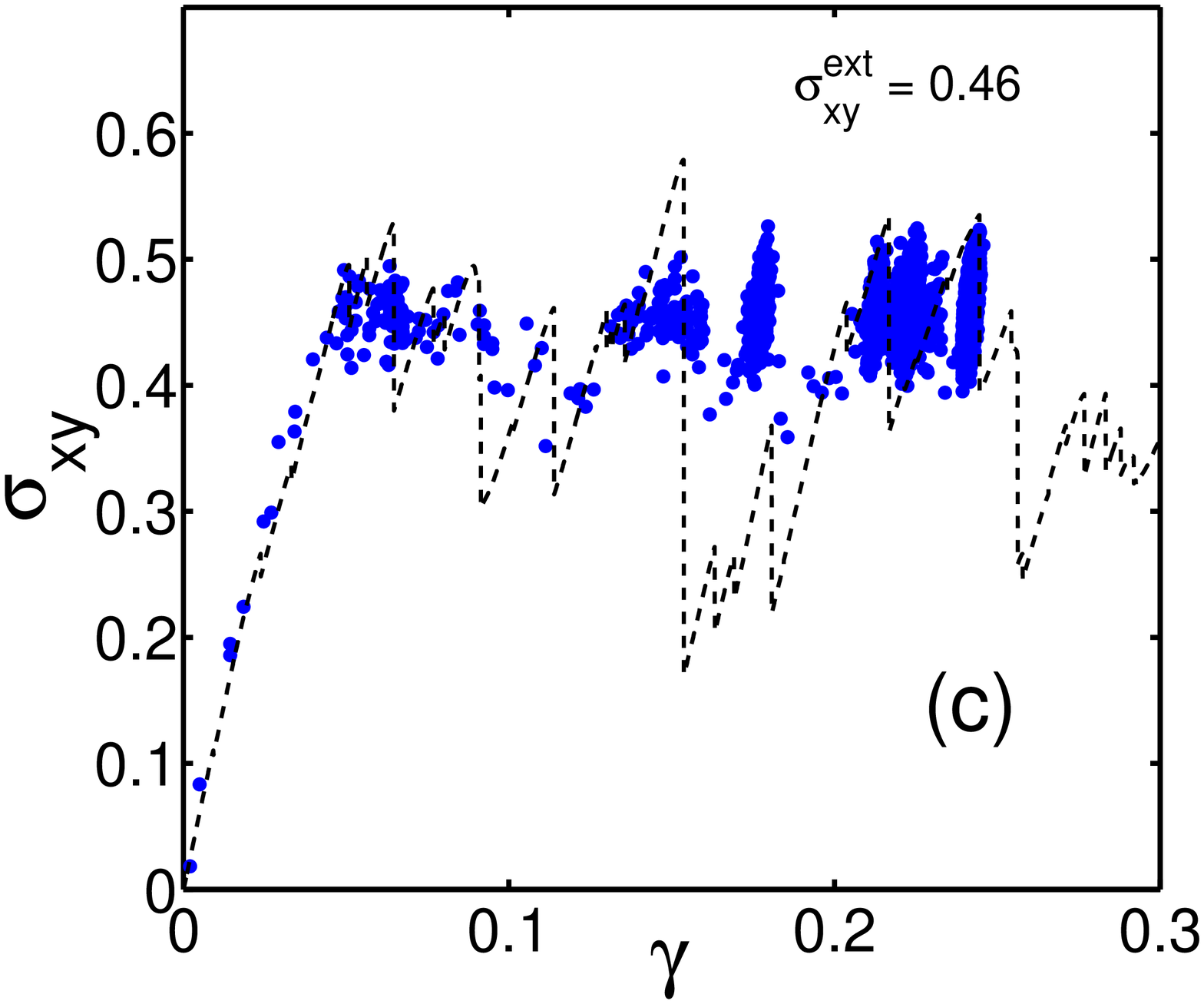}
\caption{Panel (a) : stress vs. strain in an AQS protocol that forces the system to follow the lowest energy path.
Panel (b): the Monte Carlo protocol with $\sigma^{ext}_{xy}=0.45$ and $T=0.005$. Panel (c): the Monte Carlo protocol with $\sigma^{ext}_{xy}=0.46$ and $T=0.005$.}
\label{fig3}
\end{figure}

It remains therefore to understand what {\it is} then the phenomenon of yield.  It is very helpful to consider, alongside
the stress-controlled Monte Carlo simulations, also an athermal quasistatic strain-control protocol in which the strain
is increased in steps, forcing the system to return all the time to mechanical equilibrium at $T=0$. An example of the stress-strain
dependence in such a ``AQS" protocol is shown in panel (a) of Fig.~\ref{fig3}. The elastic increases in stress are punctuated by plastic
instabilities with sharp decreases in stress (and energy). After every plastic instability the system is forced to return to mechanical
equilibrium using energy minimization. Thus the AQS protocol describes a deformation along the minimal energy path of a very complex energy landscape that is being distorted by the external strain. The system follows the series of saddle node bifurcations that are caused
by this distortion keeping itself always at the lowest possible energy. The saddle bifurcations result in shifting the system
from one local minimum of the landscape to another. Panels (b) and (c) show that the Monte Carlo stress-controlled
protocol at finite temperature follows essentially the same path. Imposing an external stress the system attempts to balance it
by increasing its internal stress, making stops in the metastable stress-strain values that correspond to the deepest basins
of the energy landscape which are exposed by the AQS plastic drops.  In other words, the process of ``yield" is neither statistical
nor random, it follows carefully the easiest path on the energy landscape.

This picture indicates that in fact there is no single ``yield-stress" but rather a series of such, and we can order them as
$\sigma_{_{Y}}^{(1)}$, $\sigma_{_{Y}}^{(2)}$ etc. in order of increasing heights of the stress peaks in the first panel of Fig.\ref{fig3}.
Of course, temperature fluctuations can cause
``glitches"  in which a the system may choose a sub-leading local minimum from which the path may deviate from the AQS analog.
Nevertheless the finite temperature path is crucially sensitive to the topography of the energy landscape greater basins.

\begin{figure}
\vskip 0.5 cm
\epsfig{width=.47\textwidth,file=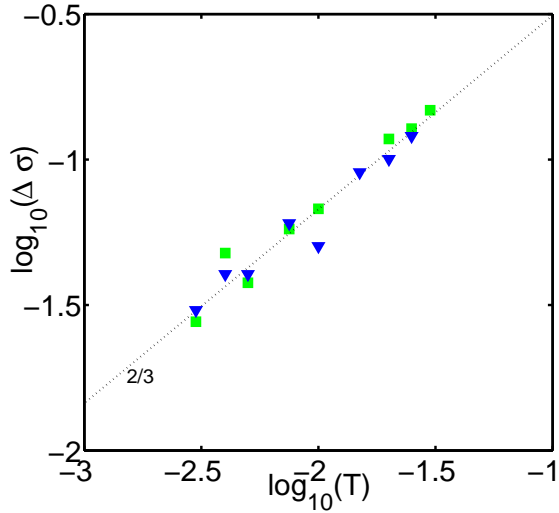}
\caption{Temperature dependence of $\Delta \sigma_{_{Y}}^{(1)}(T)$ for the first yield (corresponding to the first peak
in the AQS strain vs. stress curve), for two different realizations (squares/triangles). The slope in this log-log plot agrees well with the scaling exponents 2/3 of Eq. (\ref{sclaw}).}
\label{fig4}
\end{figure}

As additional demonstration that the stress-controlled yield is dominated by the underlying topography we turn now to the
temperature dependence of the observed yield stress $\sigma_{_{Y}}^{(n)}(T)$. We expect that if the yield follows the same
saddle-node instabilities that are so well known from AQS protocols, this must be reflected in the scaling law of
$\Delta \sigma_{_{Y}}^{(n)}(T)\equiv \sigma_{_{Y}}^{(n)}(T=0)-\sigma_{_{Y}}^{(n)}(T)$. The scaling law is easily predicted based on our knowledge that during a saddle node
bifurcation as $\gamma\to \gamma_P$ the critical eigenvalue $\lambda_P$ of the Hessian matrix goes to zero like $\sqrt{\gamma_P -\gamma}$ \cite{10KLLP,10KLP}. To find the height of the energy barrier we can expand the energy in the direction of the softening mode to obtain
\begin{equation}
U(S)=U_{\rm min} +\frac{1}{2}\lambda_P(\gamma) S^2+\frac{1}{6}K S^3+\dots.
\label{anharm}
\end{equation}
The reducing maximum is located at value of $S$ for which $dU/dS=0$ or $S\sim \sqrt{\lambda_P}$. From here
we find immediately that the energy barrier reduces like $\Delta U \sim (\sqrt{\lambda_P})^3 \sim [\gamma_P -\gamma]^{3/2}$.
At $T=0$ the barrier cannot be overcome until $\Delta U=0$. But for finite $T$ we expect that yield can occur when $T\simeq \Delta U$.
We thus conclude that we expect a scaling law (taking $\Delta \sigma\sim \Delta \gamma$):
\begin{equation}
\Delta \sigma_{_{Y}}^{(n)}(T)=C T^{2/3} \ .
\label{sclaw}
\end{equation}
This law is expected to hold {\em every time} that the external stress in the Monte Carlo stress-control protocol approaches a value
of $\sigma_{_{Y}}^{(n)}$, which can be associated with some higher
peak in the stress vs. strain graph of the AQS strain control protocol, Fig. \ref{fig3} panel (a). Indeed, in Fig.~\ref{fig4} we compare
the scaling law (\ref{sclaw}) to the measured yield stress for different temperatures associated with overcoming the first peak, i.e.
for $\Delta \sigma_{_{Y}}^{(1)}(T)$  .
Of course, the same scaling law is expected to occur close to yield for all the critical values $\sigma_{_{Y}}^{(n)}$, including the
highest which is the classically considered ``yield stress".

In general, the number and structure of those stress drops and
energy landscape basins corresponding to
$\sigma_{_{Y}}^{(1)} ... \sigma_{_{Y}}^{(n)}$,
depends on the glass preparation protocol, e.g. quench rate,
and on system size. For slow quench rates and for bigger system
one can find that the first stress-drop  $\sigma_{_{Y}}^{(1)}$
also corresponds to the highest peak, and to the final collapse
of the sample, and thus uniting the two definitions for yielding.

This $2/3$ scaling law is  also similar to the earlier reported strain-controlled  \cite{13DJHP} and experimental results \cite{05JS}, supporting the mechanical nature of this yielding phenomenon and its relation to the saddle-node bifurcations of the energy landscape.

In summary, we have shown here that stress-controlled yield in amorphous solids is akin to strain-controlled AQS protocols in the sense
that the yielding process follows (at low enough temperatures) the minimal path to increased strain, along the ``bottom" of the
energy landscape seeking the lowest saddle nodes that open up for increased strain. The AQS analog has no ``yield" since one forces the system to return to mechanical equilibrium after every plastic instability. The stress-controlled path does not equilibrate except when the internal
stress can balance the external stress. Nevertheless the stress-controlled path visits for a while the lowest available states until
it finds the cheapest escape route (see Fig.~\ref{fig3}). Of course, increasing the temperature can allow the system to escape over higher barriers and thus
to open a variance with the AQS path. Nevertheless this does not change the conclusion that the stress-controlled yield is mechanical
in nature rather than a thermal melting, either real or fictive. This conclusion is strengthened by our explicit demonstration that the
relative number of negative eigenvalues of the Hessian matrix does not change as a function of external stress even as a yield point
is approached. Finally, a clear demonstration that the yield point is close to the underlying AQS saddle-node bifurcation is
provided by the scaling law Eq. (\ref{sclaw}) which is found to be observed very accurately in our simulations. Applications
of these ideas to experimental system will be reported elsewhere.

\acknowledgments
This work had been supported in part by an ``ideas" grant STANPAS of the ERC. We thank Eran Bouchbinder for some
very useful discussions.


\end{document}